\newif\ifarxiv
\arxivtrue 

\ifarxiv
    \documentclass{article}
    \usepackage{arxiv}
\else
    \documentclass[review,11pt]{elsarticle}
    \usepackage{times}
    \usepackage{latexsym}
\fi

\usepackage{microtype}

\usepackage{amsmath,amssymb,amsfonts}
\usepackage{algorithmic}
\usepackage{graphicx}
\usepackage{textcomp}

\usepackage{tabularx}
\usepackage{tabulary}
\usepackage{makecell}
\usepackage{multirow}
\usepackage{subcaption}
\usepackage{booktabs}
\usepackage{threeparttable}
\usepackage{lscape}
\usepackage{afterpage}

\usepackage{fontawesome}
\usepackage{pifont}
\newcommand{\cmark}{\ding{51}}%
\newcommand{\xmark}{\ding{55}}%

\let\labelindent\relax
\usepackage[inline, shortlabels]{enumitem}

\ifarxiv
    \usepackage[natbib, bibencoding=utf8, citestyle=numeric, bibstyle=ieee, maxbibnames=999, maxcitenames=2, mincitenames=1, sortcites]{biblatex}
    \bibliography{refs.bib}
    
\fi

\newcommand{\eg}{e.\,g., }
\newcommand{\ie}{i.\,e., }
\newcommand{\wrt}{w.\,r.\,t.\ }

\newcommand{\wavxlsr}{\textsc{w2v2-large-xlsr}}
\newcommand{\wav}{\textsc{wav2vec2.0}}

\usepackage[acronym, shortcuts, nohypertypes={acronym}]{glossaries}
\newacronym{AI}{AI}{artificial intelligence}
\newacronym{ASR}{ASR}{automatic speech recognition}
\newacronym{compare}{\textsc{ComParE}}{Interspeech Computational Paralinguistics ChallengE}
\newacronym{COVYT}{COVYT}{Coronavirus YouTube and TikTok}
\newacronym{COVID}{COVID-19}{corona\-virus disease}
\newacronym{egemaps}{\textsc{eGeMAPS}}{extended Geneva minimalistic acoustic parameter set}
\newacronym{ML}{ML}{machine learning}
\newacronym{PCR}{RT-PCR}{reverse transcription polymerase chain reaction}
\newacronym{RBF}{RBF}{radial basis function}
\newacronym{SHAP}{SHAP}{SHapley Additive exPlanations}
\newacronym{SVM}{SVM}{support vector machine}
\newacronym{UAR}{UAR}{unweighted average recall}
\newacronym{VAD}{VAD}{voice activity detection}
\newacronym{WHO}{WHO}{world health organisation}
\newacronym{DL}{DL}{deep learning}
\usepackage[capitalise, nameinlink, noabbrev]{cleveref}

\newcommand{\covytabstract}{
More than two years after its outbreak, the COVID-19 pandemic continues to plague medical systems around the world, putting a strain on scarce resources, and claiming human lives.
From the very beginning, various AI-based COVID-19 detection and monitoring tools have been pursued in an attempt to stem the tide of infections through timely diagnosis.
In particular, computer audition has been suggested as a non-invasive, cost-efficient, and eco-friendly alternative for detecting COVID-19 infections through vocal sounds.
However, like all AI methods, also computer audition is heavily dependent on the quantity and quality of available data, and large-scale COVID-19 sound datasets are difficult to acquire -- amongst other reasons -- due to the sensitive nature of such data. 
To that end, we introduce the COVYT dataset -- a novel COVID-19 dataset collected from public sources containing more than 8 hours of speech from 65 speakers. As compared to other existing COVID-19 sound datasets, the unique feature of the COVYT dataset is that it comprises both COVID-19 positive and negative samples from all 65 speakers. 
We analyse the acoustic manifestation of COVID-19 on the basis of these perfectly speaker characteristic balanced `in-the-wild' data using interpretable audio descriptors, and investigate several classification scenarios that shed light into proper partitioning strategies for a fair speech-based COVID-19 detection.
}

\ifarxiv
    \title{COVYT: Introducing the Coronavirus YouTube and TikTok speech dataset featuring the same speakers with and without infection
    }
    
    \renewcommand{\shorttitle}{
    COVYT: Coronavirus YouTube and TikTok speech dataset
    }
    \author{
        Andreas Triantafyllopoulos\textsuperscript{1}, Anastasia Semertzidou\textsuperscript{1}, \textbf{Meishu Song\textsuperscript{1}}, \textbf{Florian B. Pokorny\textsuperscript{1, 2}}, \textbf{Bj\"{o}rn W. Schuller\textsuperscript{1,3}}\\
        \textsuperscript{1} Chair of Embedded Intelligence for Health Care and Wellbeing, University of Augsburg, Augsburg, Germany\\
        \textsuperscript{2} Division of Phoniatrics and the Division of
        Physiology, Medical University of Graz, Graz, Germany\\
        \textsuperscript{3} Group for Language, Audio, \& Music, Imperial College, London, UK\\
    }
    \renewcommand{\headeright}{Preprint}
    \renewcommand{\undertitle}{}
    
\begin{document}
    \maketitle
    \begin{abstract}
        \covytabstract
    \end{abstract}
    \keywords{COVID-19 \and speech dataset \and speech pathology \and computer audition \and disease detection \and machine learning}
\else
    \journal{Journal of Biomedical Informatics}
    \begin{frontmatter}

    \title{
    COVYT: Introducing the Coronavirus YouTube and TikTok speech dataset featuring the same speakers with and without infection
    }
    
    \author[inst1]{Andreas Triantafyllopoulos}
    \author[inst1]{Anastasia Semertzidou}
    \author[inst1]{Meishu Song}
    \author[inst1,inst2]{Florian B. Pokorny}
    \author[inst1,inst3]{Bj\"{o}rn W. Schuller}
    \affiliation[inst1]{
        organization={Chair of Embedded Intelligence for Health Care and Wellbeing,\\University of Augsburg},
        city={Augsburg},
        country={Germany}
    } 
    \affiliation[inst2]{
        organization={Division of Phoniatrics and the Division of
    Physiology, \\Medical University of Graz},
        city={Graz},
        country={Austria}
    }
    \affiliation[inst3]{
        organization={Group for Language, Audio, \& Music, \\Imperial College},
        city={London},
        country={United Kingdom}
    }
    
    \begin{abstract}
        \covytabstract
    \end{abstract}

    
    \begin{highlights}
    \item Introduced a new dataset for COVID-19 detection from free speech signals including a speaker baseline before the time of infection
    \item Learnt features outperform expert-based ones: 74.2\% vs 68.6\% UAR
    \item Spectral flux and shimmer most affected by infection amongst expert-based features
    \item Improper partitioning can lead to a dramatic over-estimation of model performance: 95.7\% vs 74.2\% UAR
    \end{highlights}
    
    
    \end{frontmatter}
\fi

\section{Introduction}
\label{sec:introduction}
In March, 2020, the \ac{WHO} has categorised the novel \ac{COVID} as a \emph{pandemic}, \ie a disease characterised by worldwide spread. 
Following this characterisation, and the immense accompanying strain on healthcare systems, several countries have taken a series of protective measures, including testing, mask mandates, movement restrictions, and vaccination campaigns -- all in an attempt to stem the devastating effects of the virus.
Today, more than two years after the outbreak of \ac{COVID}, the world is still dealing with its repercussions.
As of 1 May 2022, the \ac{WHO} has documented more than 510 million cases. A substantial number of cases was only recorded in the first quarter of 2022 due to the recent surge of the Omicron variant of \ac{COVID}.

Widespread testing is a cornerstone of the response against \ac{COVID}.
It informs public health agencies about the extent of virus spread in the community, enables the detection of new, potentially dangerous, variants, and helps citizens protect themselves and those around them by seeking timely medical assistance and self-isolating.
Currently, \ac{PCR} and rapid antigen tests dominate the testing strategies used to identify \ac{COVID} positive cases.

Recently, a plethora of \ac{AI} tools have been proposed for the automatic detection of \ac{COVID}\footnote{The term `detection' is typically used as an indication of presence of a specific event (\ie sound event detection). In our work, we are performing a classification of samples into `negative' and `positive'. Nevertheless, we use the two terms interchangeably, as detection is widely used in medical testing.}; the main justification in their favour are their significantly lower costs, their eco-friendly nature, and their potential to be deployed at a vastly larger scale.
Suggested tools analyse different types of bio-signals to make their prediction, ranging from CT-scans~\citep{wang2020covid, shah2021diagnosis, RAJAMANI2021103816}, to heart rate signals~\citep{liu2022fitbeat}, to vocal~\citep{hecker2021speaking, nessiem2021detecting, schuller2021covid, deshpande2022ai, dang2022covid} and breathing~\citep{CHEN2022104078} sounds.
So far, none of them has received medical certification and is, thus, not part of any official testing strategy mainly due to a lower accuracy as compared to standard test approaches.
However, as those tools become increasingly more sophisticated, and the virus seems to gradually transition to an endemic stage requiring less stringent monitoring, they can nicely complement the arsenal of \ac{COVID} detection mechanisms at the disposal of authorities and individuals alike.

The automatic analysis of CT-scans by means of computer vision techniques has shown much promise in detecting \ac{COVID} infection, with accuracies reaching over 90\,\% in some studies~\citep{shi2020review}; yet the major downside is that this approach requires the use of sophisticated medical equipment (computer tomographs) and the (suspected) patient to visit a medical facility -- both aspects hamper a large-scale applicability.
In contrast, heart rate and/or vocal signals can be easily obtained using everyday sensors, such as wristbands and/or smartphones, and thereby, provide a useful basis for an AI-based \ac{COVID} detection in a large group of people without a need for them to leave their homes.

In the present work, we focus on vocal sounds, in particular speech, as the bio-signal of choice for detecting and investigating the manifestation of \ac{COVID}.
\ac{COVID} as a respiratory disease suggests that acoustic information can assist in its detection: coughing, shortness of breath, and sore throat are amongst the most common reported symptoms.
Patients with mild-to-moderate symptoms frequently report dysphonia.
Accordingly, sound researchers have investigated the effectiveness of acoustic information to differentiate between patients with \ac{COVID} and controls.
A substantial amount of that work has concentrated on non-verbal sounds, such as coughing or breathing~\citep{nessiem2021detecting, coppock2021covid}. In contrast,~\citet{bartl2021voice} investigated sustained vowels. Both non-verbal sound data and sustained vowels are usually obtained through standardised procedures in controlled settings. Thereby, exactly the same type of vocal sound can be compared across patients and controls with less expectable effects of language and culture as compared to speech. However, a recorded sequence of speech covers a broader range of language-inherent sounds and sound transitions and might, thus, contain more potentially relevant acoustic information for distinguishing between individuals with and without a \ac{COVID} infection. Moreover, providing a speech sample usually represents a more natural setting for people than following the instruction to cough, breath, or produce a sustained vowel.

\ac{AI}-based digital health tools are often criticised for making unrealistic assumptions that limit their applicability in real-world applications.
Recently, \citet{coppock2021covid} provided an overview of seven specific criticisms on audio-based \ac{COVID} detection -- denominated as seven `grains of salt':
\begin{enumerate}
    \item Just investigating \ac{COVID} vs healthy condition (neg\-lecting other diseases),
    \item Presence of (confounding) background noise,
    \item Subject knowledge of infection status which potentially impacts vocal expression (\eg through emotion),
    \item Questionable validity of (self-reported) COVID-19 status,
    \item Lack of sound data and code availability,
    \item Ignorance of demographic characteristics,
    \item Lack of speaker-disjoint experiments.
\end{enumerate}

In the present work, we introduce the \ac{COVYT} speech dataset, which attempts to mitigate several of those issues.
Using an easily scalable collection and pre-processing protocol allowing to make data from social media channels scientifically exploitable, we provide a unique multilingual dataset for investigating COVID-19 detection from free speech that features \ac{COVID} positive and negative speech samples from exactly the same speakers. 
Thereby, we guarantee for minimal bias potentially introduced by imbalances in intrinsic speaker characteristics, such as gender, age, and language.
In addition, the presence of both positive and negative samples from the same speakers allows us to explore personalisation approaches which disentangle the effects of infection from individual voice characteristics that can confound analysis~\citep{dang2022covid}.
We present dataset baselines with respect to (i) the manifestation of a \ac{COVID} infection in different acoustic descriptors and (ii) automatic \ac{COVID} detection in various scenarios addressing several factors that influence model performance.
The \ac{COVYT} dataset as well as the code for our machine learning experiments are publicly available to facilitate reproducibility and to motivate further research comparable to the provided baselines\footnote{\#\# Links will be added upon manuscript acceptance.}.

The remainder of this work is organised as follows.
In \cref{sec:covyt}, we introduce our \ac{COVYT} speech dataset and present related data collection and pre-processing protocols, statistics, and partitioning.
In \cref{sec:related}, we compare the \ac{COVYT} speech dataset to other relevant currently existing vocal sound datasets for \ac{COVID} detection.
\cref{sec:evaluation} then reveals our dataset baselines, while \cref{sec:discussion} positions the \ac{COVYT} dataset and our findings with respect to previous research and \cref{subsec:acousticAnalysis} discusses the strengths and limitations of our work, connecting it to the above mentioned seven `grains of salt'.
We conclude our work in \cref{sec:conclusion}.

\section{COVYT dataset}
\label{sec:covyt}

\subsection{Data collection}
\label{sec:collection}

As the pandemic is a global phenomenon that has dominated the public's attention since its very beginning, people -- and in particular celebrities -- have often `announced' their positive results on social media.
Some of those cases, like that of the former US president Donald Trump Jr., receive considerable media attention due to the nature and position of the person affected.
It became common for news outlets to run features using footage of well-known people discussing their experiences on having (had) a \ac{COVID} infection; or even the celebrities spread footage by themselves through their private channels.
Respective footage is typically recorded in the days following a positive \ac{COVID} test, when subjects are required to stay in quarantine. Here, the \ac{COVID} status labels are self-reported and cannot be officially verified.

The data collection phase for the \ac{COVYT} speech dataset took place between November 2020 and November 2021. 
During that time, two popular media platforms are combed for appropriate material, namely YouTube and TikTok. Different data collection protocols for positive cases are utilised for each platform.
A targeted search is performed on YouTube, where high-profile cases, \eg actors, politicians, celebrities, etc. that come to the authors' attention through the media, are intentionally looked for.
In contrast, a global search is performed on TikTok by using keywords like ``COVID`` or ``symptoms``.
In both scenarios, when finding an eligible \ac{COVID} positive example, we search for uploads of the same speakers preceding the date of infection to serve as negative examples. 
In case we easily find more than one positive or negative examples of a speaker, all clips are included. 
Different protocols serve to mitigate potential biases in our data collection process by incorporating diverse recording scenarios, ranging from high-quality, professional interviews to homemade smartphone videos. 
In all identified clips, the speakers were audio-video recorded or recorded themselves, while talking, \eg during an interview, a public speech, a narrative, or a `story'.
As most clips were released by speakers to explicitly inform the community about their infection status, we are able to harness them for acoustic analysis and model training.
Recordings during the \ac{COVID} positive state were taken mostly indoors -- at home, a hospital, a TV/ Radio studio, or a press conference room -- as subjects had to be under quarantine.
To avoid undesired interference, only videos with minimal background noise or quiet music, \eg as known from news reports and media covers, are accepted for this work. 
In total, we include $185$ videos -- $89$ positive and $96$ negative examples. 
Of these, $83$ just contain a single speaker, whereas the rest contain multiple ones -- requiring additional processing to extract the utterances of the target speaker.


\subsection{Data preparation}
\label{sec:preparation}

First of all, we download all identified videos in .MP4 format using freely available tools. 
We then manually annotate the acoustic environment of each video according to recording location -- `indoor' vs `outdoor' -- and recording setting. 
For recording setting, we distinguish between:
\begin{enumerate*}
    \item speeches (or press-releases), which are longer recordings, where the target speaker releases a statement in front of staged, professional-grade cameras;
    \item interviews, where the target speaker is part of a (present or online) conversation; and 
    \item self-recordings, where the target speaker uses his or her own smartphone to make a short video (usually a social-media-style `story').
\end{enumerate*}
Following this annotation, we extract the audio streams and store them as .WAV files in format 16\,kHz, 16\,bit, single-channel, PCM.
The clips range in duration from $10$\,s to $1$\,h\,$11$\,m -- some clips contain long monologues of the target speaker.
Therefore, we segment all clips into single utterances for further processing.
We choose a semi-automatic segmentation approach conservative in the amount of utterances it keeps.
We employ an recurrent neural network (RNN)-based \ac{VAD} model~\citep{hagerer2017enhancing} as the initial stage, followed by a manual verification stage using ELAN\footnote{https://archive.mpi.nl/tla/elan}~\citep{wittenburg2006elan}.
Utterances that do not exclusively contain speech of the target speaker or samples that contain music, or background noise are excluded.

\subsection{Facts and figures}
\label{sec:facts}

An overview of dataset statistics is given in \cref{tab:data}.
The \ac{COVYT} dataset contains $10\,413$ utterances with a total duration of $8$\,h\,$15$\,m of speech from $65$ speakers -- $25$ females and $40$ males -- at ages ranging from $23$ to $74$\,years at time of infection (mean $=46$\,years $\pm$ $13$\,years standard deviation). Each speaker has a median number of $113$\,utterances -- $27$ \ac{COVID} positive and $66$ \ac{COVID} negative examples. Henceforth, the respective point in time at which the speakers had a COVID-19 infection is referred to as T+; the point in time at which they did not have a COVID-19 infection is referred to as T--. Speakers are celebrities of various domains: actors, athletes, journalists, models, musicians, politicians,  presenters, reporters, singers, and writers.
Moreover, the dataset covers $9$ different languages, namely Chinese (CN), English (EN), French (FR), German (DE), Greek (GR), Polish (PL), Portuguese (PT), Slovakian (SK), and Spanish (ES), with a majority of English speakers ($40$) followed by Greek speakers ($14$).
The number of speakers and utterances is given in detail in \cref{tab:data}.
With regard to location and setting, most clips were recorded indoors rather than outdoors (T--: 92 vs 4; T+: 83 vs 6); most clips were recorded in an interview setting (T--: 55; T+: 27), followed by a speech/press-release setting (T--: 29; T+: 6), and, lastly, by a self-recording setting (T--: 12; T+: 56).
Naturally, the chance to give a speech or an interview decreases following a \ac{COVID} diagnosis, as speakers have to self-quarantine or are hospitalised for a certain time.
Interviews at T+ took place through online teleconferencing, and speeches were made in front of staged cameras (presumably without the presence of reporters or assistants due to quarantine restrictions).

    \begin{table}
    \centering
    \scriptsize
    \caption{
    Overview of dataset statistics.
    We show the language-wise and total number of (\#) speakers as well as \# utterances and utterance duration for the respective points in time at which the speakers had a \ac{COVID} infection (T+) and the respective points in time at which the speakers did not have a \ac{COVID} infection (T--). Gender-wise numbers and durations are given in parentheses in order male / female. 
    }
    \label{tab:data}
    \begin{threeparttable}
    \begin{tabular}{r|l|lr|lr}
            \toprule
            & & \multicolumn{2}{c|}{\textbf{T+}} & \multicolumn{2}{c}{\textbf{T--}}\\
            \textbf{Language} & \textbf{\# Speakers}  & \textbf{\# Utterances} & \textbf{Duration (hh:mm:ss)}  & \textbf{\# Utterances} & \textbf{Duration (hh:mm:ss)} \\
    \midrule
    Chinese &     \phantom02 (0 / 2) &      \phantom{0\,}112 (0 / 112) &   00:03:43 (00:00:00 / 00:03:43) &        \phantom{0\,0}28 (0 / 28) &   00:00:53 (00:00:00 / 00:00:53) \\
    English &  40 (28 / 12) &  2\,454 (1\,988 / 466) &   01:59:45 (01:36:00 / 00:23:45) &  4\,686 (4\,068 / 618) &   03:45:54 (03:12:52 / 00:33:03) \\
    French &     \phantom02 (2 / 0) &        \phantom{0\,0}84 (84 / 0) &   00:04:10 (00:04:10 / 00:00:00) &        \phantom{0\,0}96 (96 / 0) &   00:03:04 (00:03:04 / 00:00:00) \\
    German &     \phantom02 (1 / 1) &       \phantom{0\,0}27 (16 / 11) &   00:01:26 (00:00:48 / 00:00:38) &       \phantom{0\,0}85 (65 / 20) &   00:04:28 (00:02:59 / 00:01:29) \\
    Greek &    14 (5 / 9) &    \phantom{0\,}544 (187 / 357) &   00:29:28 (00:08:33 / 00:20:55) &    \phantom{0\,}869 (179 / 690) &   00:43:27 (00:08:46 / 00:34:41) \\
    Polish$^*$ &     \phantom01 (1 / 0) &        \phantom{0\,0}23 (23 / 0) &   00:02:01 (00:02:01 / 00:00:00) &      \phantom{0\,}254 (254 / 0) &   00:11:59 (00:11:59 / 00:00:00) \\
    Portuguese &     \phantom01 (1 / 0) &        \phantom{0\,0}21 (21 / 0) &   00:00:50 (00:00:50 / 00:00:00) &      \phantom{0\,}118 (118 / 0) &   00:05:33 (00:05:33 / 00:00:00) \\
    Slovakian &     \phantom01 ( 1 / 0) &        \phantom{0\,0}54 (54 / 0) &   00:02:54 (00:02:54 / 00:00:00) &      \phantom{0\,}273 (273 / 0) &   00:12:41 (00:12:41 / 00:00:00) \\
    Spanish &     \phantom02 (1 / 1) &       \phantom{0\,0}35 (18 / 17) &   00:01:02 (00:00:29 / 00:00:33) &     \phantom{0\,}650 (632 / 18) &   00:23:35 (00:22:32 / 00:01:03) \\
    \midrule
    \textbf{Total} & \textbf{65 (40 / 25)} & \textbf{3\,354 (2\,391 / 963)} & \textbf{02:45:18 (01:55:45 / 00:49:33)} & \textbf{7\,059 (5\,685 / 1374)} & \textbf{05:31:35 (04:20:26 / 01:11:09)}\\
    \bottomrule
    \end{tabular}
    \begin{tablenotes}
        \item[$*$] A portion of the negative utterances of this speaker is actually in English; however, we consider this to have a negligible effect on our analysis.
    \end{tablenotes}
    \end{threeparttable}
    \end{table}
    

\subsection{Partitioning}
\label{subsec:partitioning}

Proper partitioning is a crucial aspect of any dataset if used for \ac{ML} purposes, as the data must be split in a way that allows for building well-performing models, but also enables a fair evaluation.
Taking into account the relatively small size of the \ac{COVYT} dataset, with a total of 10\,413 samples, we opt for a cross-validation scheme, for which we provide training/development/testing folds.
Given that we also aim to investigate different scenarios of \ac{COVID} detection, we introduce four different partitioning strategies, each targeted to a different aspect of interest.

\begin{enumerate}
    \item \textbf{Speaker-disjoint partitioning:} As discussed in \citet{coppock2021covid}, a major limitation of several existing \ac{COVID} datasets is their lack of subject-disjoint evaluation. 
    If data from the same speaker is involved in both training and testing, performance will be most probably higher because the model might re-identify the speaker identity instead of performing the actual target task.
    Individual voice characteristics often negatively impact generalisation; having the same speaker in the training and testing partitions results in obtaining over-optimistic performance scores.
    This particular partitioning scheme is implemented by randomly splitting the speakers into $3$ disjoint groups (${G_1, G_2, G_3}$), and subsequently taking all possible permutations of this set, resulting in $6$ folds (the permutations are obtained by training on $G_1$, validating on $G_2$, testing on $G_3$, then training on $G_2$, validating on $G_1$, testing on $G_3$ -- so each fold is train/validated/tested on twice).
    \item \textbf{Speaker-inclusive partitioning:} This partitioning is intentionally introduced to quantify the effect of having negative and positive data of the same speakers in training and testing.
    The data of all files from each speaker is randomly split into three partitions; then, the partitions are permuted to obtain $6$ folds in similar fashion as before.
    \item \textbf{File-disjoint partitioning:} With this partitioning strategy, we want to investigate whether a speaker-inclusive, but file-disjoint partitioning would also yield a disproportionately good performance. 
    In particular, we ensure that the same speaker is present in the test and either the training or development partitions, but the same file is not.
    This enables us to quantify whether it is indeed the individual speaker characteristics that cause models to overperform in the speaker-inclusive scenario, or whether it is simply the side-effect of having an identical recording (which contains data from the same acoustic conditions).
    Given that for most speakers the \ac{COVYT} dataset contains only one original clip per state (\ac{COVID} negative or \ac{COVID} positive), we randomly split speakers into two groups, $G_1$ and $G_2$.
    We then create two test sets, $T_1$ and $T_2$, with $T_1$ containing all $T+$ instances of $G_1$ and all $T--$ instances of $G_2$, and $T_2$ in contrast with all $T--$ instances from $G_1$ and all $T+$ instances of $G_2$.
    This ensures that each test set contains files from all speakers but in only one of the two conditions (\ac{COVID} negative or \ac{COVID} positive), thus, ensuring that files are disjoint (as each recording contains the speaker in only one condition).
    For each fold, we additionally create $2$ variants where the training set is randomly split into  $2$ training/development sets (this time in speaker-disjoint fashion, as most speakers have only one clip per condition), resulting in a total of $4$ folds when taking all permutations.
    \item \textbf{Language-disjoint partitioning:} For this final scheme, we split the data by taking language information into account.
    As most of our data is either English or Greek, we bundle all other languages in a separate \emph{`other'} group, thus resulting in $3$ folds.
    Once again, we utilise all permutations to create $6$ folds.
\end{enumerate}

We note that the information regarding partitioning is included in the metadata released together with the dataset, thus ensuring reproducibility of our results and a fair comparison of different scenarios.
However, not all partitioning schemes are relevant for potential real-world applications. 
We encourage authors to primarily use our \emph{speaker-disjoint} partitioning for future work.
Moreover, it is often the case that practitioners try to avoid the computational overhead of cross-validation (in particular for resource intensive \ac{DL} architectures that are currently state of the art in the field).
While we encourage the proper use of all folds to get a more robust estimation of performance and generalisation, we acknowledge the practical considerations and recommend authors that want to avoid the overhead to use the 1\textsuperscript{st} fold of the speaker-disjoint scheme ($G_1$).
To facilitate comparisons with future studies, we will also include those results in our manuscript, even though we primarily focus on our full cross-validation experiments.

\section{\ac{COVYT} vs other datasets}
\label{sec:related}
    \begin{table}[t]
        \caption{
        Comparison of the \ac{COVYT} dataset with other datasets for audio-based \ac{COVID} detection.
        }
        \label{tab:datasets}
        \centering
        \scriptsize
        
        \begin{threeparttable}
        \begin{tabular}{c|cccccccc}
            \toprule
            \textbf{Dataset} & \textbf{\makecell{\# Speakers in total ($+$)}} & \textbf{Sound type} & \textbf{\# Languages} & \textbf{Data availability} & \textbf{\ac{COVID} labels} & \textbf{$\pm$ from same speaker}\\
            \midrule
            \midrule
            \textbf{COVIDTelephone\citep{ritwik2020covid}} & 19 (10) & free speech & N/A & \cmark & level-1 & \xmark \\
            \hline
            \textbf{AI4COVID\citep{imran2020ai4covid}} & 543 (70) & coughing & N/A  & \xmark & unknown$^{*}$ & \xmark\\
            \hline
            \textbf{Coswara~\citep{sharma2020coswara, muguli21_interspeech}} & 941 (104) & \makecell{breathing\\coughing\\vowels\\digit counting} & multiple$^{**}$ & \cmark & level-1 & \xmark\\
            \hline
            \textbf{COUGHVID~\citep{orlandic2021coughvid}} & 14787 (410 & coughing & N/A & \cmark & level-1 & \xmark\\
            \hline
            \textbf{YourVoiceCounts~\citep{bartl2021voice}} & 22 (11) & \makecell{coughing\\vowels\\read speech} & 1 & \xmark & level-2 & \xmark\\
            \hline
            \textbf{CoughAgaistCovid~\citep{bagad2020cough}} & 3621 (2001) & coughing & N/A & \xmark & level-2 & \xmark\\
            \hline
            \textbf{COVID19~\citep{pinkas2020sars}} & 78 (29) & \makecell{coughing\\vowel / voiced consonant\\counting digits} & 1$^{***}$ & \xmark & level-2 & \xmark\\
            \hline
            \textbf{\makecell{YourVoiceCounts \\audEERING~\citep{hecker2021speaking}}} & 39 (19) & \makecell{coughing\\vowels\\read speech} & 1 & \xmark & \makecell{level-1/2} & \xmark\\
            \hline
            \textbf{DiCOVA~\citep{muguli21_interspeech}} & 366 (-) & \makecell{coughing\\read speech} & 8 & \faLock & level-2 & \cmark \\
            \hline
            \textbf{\makecell{Cambridge\\Longitudinal~\citep{dang2022covid}}} & 212 (106) & \makecell{breathing\\coughing\\read speech} & 8 &  \faLock & level-1 & \cmark \\
            \hline
            \textbf{COVID-19 Sounds~\citep{xia2021covid}} & 36116 (2496) & \makecell{breathing\\coughing\\read speech} & 8 & \faLock & level-1 & \cmark\\
            \midrule
            \midrule
            \textbf{COVYT} & 65 (65) & free speech & 9 & \cmark & level-1 & \cmark \\
            \bottomrule
        \end{tabular}
        \begin{tablenotes}
            \item[] \# = number of; $+$ = \ac{COVID} positive; $\pm$ = \ac{COVID} positive and \ac{COVID} negative (samples)
            \item[] \emph{level-1} self-assessment / no test required; \emph{level-2} Medically certified \ac{COVID} test result required
            \item[$*$] Not specified; however, most likely level-2 as the study was conducted in medical facilities
            \item[$**$] Speakers came mainly from India, which has 2 official and 22 recognised regional languages.
            \item[$***$] Not mentioned; presumably Hebrew as the study was conducted in Israel
       \end{tablenotes}
        \end{threeparttable}
    \end{table}

\cref{tab:datasets} provides a representative overview of currently existing \ac{COVID} sound datasets and specifies aspects that the \ac{COVYT} dataset is meant to set value on.
In particular, we focus on the type of audio content that each dataset contains.
The \ac{COVYT} dataset contains free (multilingual) speech; this is in contrast to most other listed datasets that primarily focus on breathing and coughing sounds, on sustained vowels, or on read standard texts.
Besides some advantages of free speech (please also see~\cref{sec:introduction}; coverage of several language-inherent sounds, natural recording setting), our motivation to use this type of sound data is a pragmatic one: Free speech is most easily available on our data sources.
The amount of different languages contained in a dataset and their distribution are mostly relevant for those including linguistic sound types, such as speech or sustained vowels; however, it also serves as a proxy for demographic diversity which is an important consideration for making datasets fair and audio-based \ac{COVID} detection applicable across different countries. This is why we make the \ac{COVYT} dataset as diverse as possible and provide different partitioning strategies including a language-disjoint one.
Data availability is in turn crucial for ensuring study reproducibility and transparency of research findings, and also to foster new advances from researchers without access to own datasets. 
Here, the \ac{COVYT} dataset enjoys an advantage over other datasets, as it is sourced entirely from the public domain.

This leads to the issue of label reliability.
Relying on self-reported labels and crowdsourced data holds lots of potential for scaling the size of a dataset. However, it comes with the danger of erroneous labels.
In contrast, manually collected datasets with strict medical protocols, where the label is verified through proper medical examination and further audited by a medical practitioner is the gold standard for digital health applications, but is harder to scale due to the amount of resources it requires.
Our approach lies somewhere in between.
While technically, we still rely on self-reported labels, we make use of the scrutiny that celebrities come under from the press, which renders a fake \ac{COVID} report rather unlikely (though not impossible).
To roughly distinguish between different quality types of labelling protocols, we adopt a $2$-level system, which ranks datasets according to whether they solely rely on self-reported labels (\emph{level-1}), or whether a medically certified test result was required (\emph{level-2}).
The \ac{COVYT} dataset falls under the \emph{level-1} category.

Finally, we include another indicator in our comparison, namely whether a dataset contains samples of the same speaker with and without infection. We consider this as an important aspect towards minimising potential biases caused by imbalances in specific speaker characteristics, such as gender, age, or any other intrinsic anatomical/voice-physiological properties. Investigating exactly the same voice with vs without infection across many speakers, respectively, increases the chance to effectively identify disease-related phenomena. 
Moreover, recent findings from the \textit{Cambridge Longitudinal} dataset~\citep{dang2022covid} suggest that there are individual effects in the manifestation of \ac{COVID} in the voice (in their case, non-linguistic vocalisations). 
To the best of our knowledge, the \ac{COVYT} dataset is the only dataset alongside \textit{Cambridge Longitudinal} and \textit{COVID-19 Sounds} (both variants of the same data) to fulfill the criterion of having \ac{COVID} positive and negative samples from the same speakers, and, in particular, the only one to fulfill this criterion and to contain free speech samples, which enables the use of personalised \ac{ML} algorithms, which have been shown to improve performance in other speech-based tasks~\citep{triantafyllopoulos2021deep}. 
However, it has to be considered, that the recording setting (recording equipment, location, situation, speaker mood, etc.) might differ between the respective positive and negative sample of one and the same speaker, which potentially introduces systematic acoustic bias.


\section{Baseline evaluations}
\label{sec:evaluation}

In the following, we provide standard solutions for answering two main research questions (RQs) on the basis of \ac{COVYT} data: (\ref{subsec:acousticAnalysis}) Which speech parameters differ most between speakers at T-- vs T+? (\ref{subsec:detection}) Can T-- and T+ be automatically differentiated from speech? To this end, we derive different audio representations from the available utterances. The generated results shall serve as a benchmark for future analyses carried out on the basis of the \ac{COVYT} dataset.       

\subsection{Audio representations}
\label{subsec:audioRepresentations}

Speech processing applications typically rely on hand-crafted sets of less than 100 until several 1000 features, which attempt to holistically describe an input utterance~\citep{eyben2015geneva, schuller2013interspeech}; though, certain features allow for signal interpretation from a voice-physiological perspective.
In recent years, however, learnt representations have shown superior performance and robustness in several tasks~\citep{baevski2020wav2vec, conneau21crosslingual}, driving their adaptation from the community.
In the present study, we carry out experiments using three different audio representations:

We begin with the \ac{egemaps}~\citep{eyben2015geneva} -- a rather small set of $88$ acoustic parameters that has previously been shown to contain relevant information for the manifestation of COVID-19 in sustained vowels~\citep{bartl2021voice}.
In contrast, the \ac{compare} set is a large-scale feature set of 6\,373 acoustic parameters. As the official baseline feature set of the \ac{compare} series from 2013 until 2021~\citep{schuller2013interspeech,schuller2021interspeech}, it has been successfully used for several computer audition tasks over the last decade.
Both the \ac{egemaps} and the \ac{compare} set are extracted using the open-source toolkit openSMILE~\citep{eyben2010opensmile}.

Finally, we use learnt representations from \wavxlsr~\citep{conneau21crosslingual}, a multilingual variant of \wav~\citep{baevski2020wav2vec}.
This model was pre-trained in self-supervised fashion on a large corpus containing $53$ languages.
We thus expect this network to generalise better to our multilingual data than the vanilla \wav.
The architecture consists of $7$ convolutional neural network (CNN) feature extraction layers followed by $24$ transformer (self-attention) layers.
Here, we use the intermediate features which are extracted by the CNN layers.
These roughly correspond to $25$\,ms of audio with a stride of $20$\,ms, which we subsequently average over time to obtain the final embeddings.
We also experiment with the contextualised representations, which are extracted after the self-attention layers. 
However, we get consistently worse performance.
As our intention here was to obtain competitive baselines, we do not fine-tun {\wavxlsr} on COVYT, even though this is not to lead to substantially better performance~\citep{wagner2022dawn}.

\subsection{Acoustic analysis (RQ\,1)}
\label{subsec:acousticAnalysis}

The analysis of acoustic differences between speakers at T+ vs T-- is done on the basis of the extracted \ac{egemaps} representation, as \ac{egemaps} features generally offer interpretability from a clinical/voice-physiological perspective. To ensure equal speaker and COVID-19 status weighting, we average feature values across all utterances of a single speaker at T+ and T--, respectively. Thus, for each of the 88 \ac{egemaps} features, we produce exactly one value per speaker at T+ and one value per speaker at T--. We find that the values of the single feature at T+ and T-- are not normally distributed. Thus, we feature-wisely apply the Mann-Whitney U test and derive the effect size $r$, i.\,e., the absolute value of the correlation coefficient calculated as the $z$-value divided by the square root of the number of samples~\citep{Rosenthal94-PMO}. We finally rank the \ac{egemaps} features according to the effect size and define top features to have an $r>0.3$ (fair correlation at minimum). In addition, we report two-sided $p$-values.

\begin{figure}[t]
    \centering
    \includegraphics[width=1\textwidth]{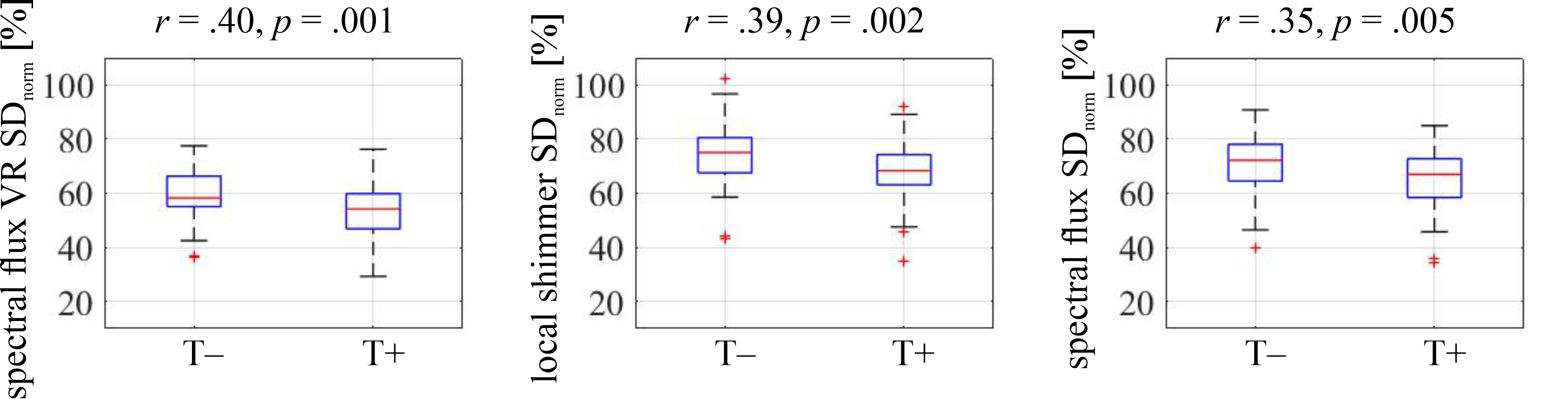}
    \caption{Comparison of speakers at their respective point in time without COVID-19 infection (T--) vs the point in time with COVID-19 infection (T+) by means of boxplots for the three identified acoustic features with a differentiation effect $r>0.3$. Boxplots are ordered according to a decreasing $r$ from left to right. Effect size $r$ and $p$-value of the Mann-Whitney U difference test are given above each boxplot. $r$ is rounded to two decimal places. $p$ is rounded to three decimal places. Outliers (red plus symbols) are defined as values more than 1.5 times the interquartile range away from the bottom or top of the respective box. SD\textsubscript{norm} = standard deviation normalised by arithmetic mean (= coefficient of variation), VR = voiced regions}
    \label{fig:boxplots}
\end{figure}

\begin{figure}[t]
    \centering
    \includegraphics[width=1\textwidth]{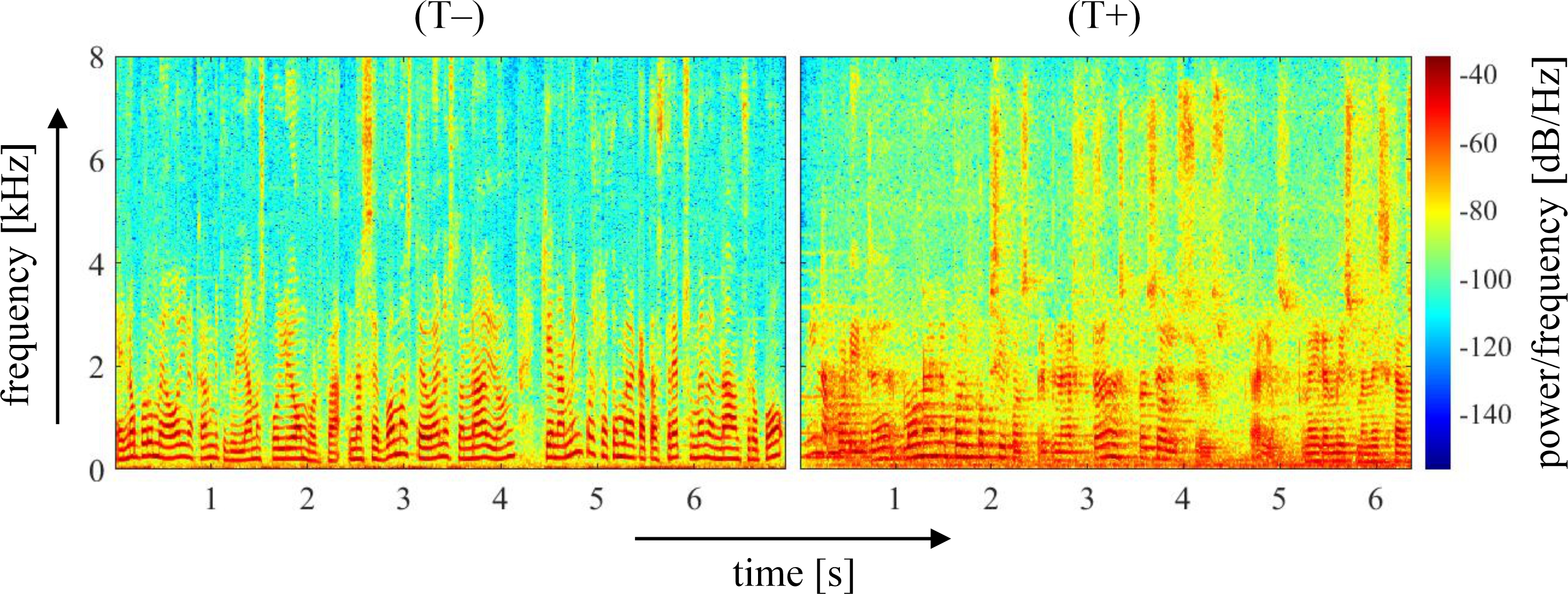}
    \caption{
    Spectrograms of utterances produced by one and the same speaker (gender: female, nationality: Greek, age at T+: 43) at a point in time without COVID-19 infection (T--) and at a point in time with COVID-19 infection (T+).
    }
    \label{fig:spectrograms}
\end{figure}

We identify three top features, namely (I) the coefficient of variation of the spectral flux, i.\,e., spectral change between consecutive time frames~\citep{eyben2015geneva}, in voiced regions, (II) the coefficient of variation of local shimmer, i.\,e., change in amplitude between consecutive fundamental frequency periods~\citep{eyben2015geneva}, and (III) the coefficient of variation of the spectral flux in the entire speech segment. \cref{fig:boxplots} reveals the respective boxplots for T-- vs T+ alongside the effect sizes and $p$-values. In all three top features, T+ is characterised by a lower coefficient of variation as compared to T--. This means that there is restricted variation with regard to spectral and amplitude change within an utterance in COVID-19-related speech. \cref{fig:spectrograms} exemplarily shows the spectrograms of an utterance produced at T-- and an utterance produced at T+. Both utterances originate from the same speaker, namely the speaker with the highest average top feature (coefficient of variation of spectral flux in voiced regions) difference between T+ and T--. Obviously, the utterance produced at T+ exhibits more inharmonic overtones in voiced sounds as compared to the utterance produced at T--, which is associated with more vocal coarseness as typical for a respiratory disease. Moreover, the presented spectrogram related to T+ indeed suggests less variation of spectral change over time in voiced regions. However, this finding has to be interpreted with caution, as not only mechanisms of voice production but also room acoustics affect the recorded audio signal and, thereby, a wide range of derived acoustic features. An auditory inspection of the utterances presented in \cref{fig:spectrograms} yields that the utterance produced at T+ was obviously recorded in a room with a longer reverberation time as compared to the utterance recorded at T-- (please also see limitations discussed in \cref{subsec:strengthsWeaknesses}).  

\subsection{Automatic COVID-19 detection (RQ\,2)}
\label{subsec:detection}

In this subsection, we present our automatic COVID-19 classification performance results on all four partitioning strategies described in \cref{subsec:partitioning}.
In all cases, we perform z-score normalisation on each feature separately, computing the statistics on the training set of each fold in our cross-validation setup separately, and consequently applying them on the development and test partitions.
As a typical baseline classifier, we utilise \acp{SVM}, where we optimise the complexity parameter $C$ in [$.0001$, $.0005$, $.001$, $.005$, $.01$, $.05$, $.1$, $.5$, $1$] as well as the kernel function across the types linear, polynomial, and \ac{RBF}.
Optimisation is always done on the development partition of each fold.
We additionally utilise two distinct evaluation protocols, each highlighting a different aspect of the underlying problem.
The first is adhering to standard \ac{ML} practice.
As discussed in \cref{sec:preparation}, the downloaded clips are segmented into utterances, which we use as training/validation/testing instances in the context of a \ac{ML} pipeline.
These instances inherit the label (presence or absence of a \ac{COVID} infection) from the file they originate from.
In each fold, the model is trained on the training instances and evaluated on the test instances, resulting in one prediction value per instance.
These predictions are then evaluated against the ground truth labels over the entire test set to provide a general, instance-level performance score.
This procedure is repeated for each fold in each partitioning scheme.
The second protocol is motivated by the clinical evaluation setting for which our application is intended.
Under this perspective, the different utterances resulting from the segmentation process can be seen as repeated measurements of the same underlying variable, namely the manifestation of \ac{COVID} in the speaker's voice.
In this setting, instance-level decisions are aggregated to provide a final, holistic evaluation, which takes all utterances into account and provides a single label for each file in the test partition.
As our present focus is on providing a set of competitive baselines, we adopt the simplest possible aggregation process, namely \emph{max voting}, where the label corresponding to each file is defined as the most often-predicted label across all its utterances.
As our metric of choice we use the \ac{UAR}, which accounts for a certain label imbalance in some folds.

We present \ac{UAR} results for all partitioning scheme and audio representation combinations in \cref{table:results}.
At first glance, \wav results in the highest performance across all schemes, followed by \ac{compare}, while \ac{egemaps} is lagging further behind.
This demonstrates the power of learnt representations for \ac{COVID} detection in free speech samples.

The different partitioning schemes allow us to find the answers of two important questions:
\begin{enumerate}
    \item \textit{How important is the use of speaker-disjoint sets?} 
    The vastly superior performance of all audio representations on speaker-inclusive experiments, where \wav\ reaches an average \ac{UAR} of 98\,\% on the instance- and 96\,\% on the file-level, compared to the alternative partitioning strategies, where \wav reaches a maximum average \ac{UAR} of 71\,\% and 74\,\% on instance- and file-level evaluations, respectively, shows that this partitioning can result in a large overestimation of real-world performance and should, therefore, be avoided.
    However, the drop of performance observed when switching to file-, but not speaker-disjoint evaluations also indicates that it is not necessarily the presence of the same speaker in training and testing partitions per se that causes this overestimation, but potentially (also) the splitting of the same recording (whose utterances form repeated measurements of the same speaker in the same acoustic environment).
    Actually, the difference between our speaker-disjoint and file-disjoint partitioning schemes is negligible, showing that the speaker effect is small when taking care to avoid other confounders (\eg recording conditions, background noise, etc.). 
    Still, to be on the safe side, future work on the \ac{COVYT} dataset and comparable datasets (where applicable) should avoid the use of speaker-inclusive partitions in order to obtain accurate generalisation estimates.
    \item \textit{Is a speech-based detection of \ac{COVID} universally possible, i.\,e., across languages?}
    The evidence here is inconclusive.
    Our language-disjoint partitioning turns out to be the most challenging for the audio representations tested here, with performance dropping to 63/66\,\% even for the best-performing \wav, while resulting in random-chance performance for \ac{egemaps} (49/51\,\%).
    This is expected, as several speech-based computer audition tasks struggle with cross-language generalisation.
    Even so, the fact that the learnt representations of \wav, which has been pre-trained on several languages, still shows the highest performance is a promising sign that future `multilingual' foundation models might bridge this gap and form the basis of a universal \ac{COVID} detection model.
    Though, until that time, it is highly recommended that models are applied with special care to different languages/cultures, and (if possible) evaluated on a representative test set beforehand.
\end{enumerate}

\begin{table}[t]
    \caption{Cross-validation unweighted average recall (UAR) results for all partitioning strategies and audio representations using support vector machines (SVMs) with different features.
    }
    \label{table:results}
    \centering
    \resizebox{\columnwidth}{!}{
    \begin{tabular}{c|cc|cc|cc|cc}
    \toprule
    \textbf{Partitioning} & \multicolumn{2}{c|}{\textbf{Speaker-disjoint}}& \multicolumn{2}{c|}{\textbf{Speaker-inclusive}} & \multicolumn{2}{c|}{\textbf{File-disjoint}}& \multicolumn{2}{c}{\textbf{Language-disjoint}}\\
    \textbf{Audio representation} & \textbf{\makecell{Segment-level\\UAR [\%]}} & \textbf{\makecell{File-level\\UAR [\%]}} &  \textbf{\makecell{Segment-level\\UAR [\%]}} & \textbf{\makecell{File-level\\UAR [\%]}} & \textbf{\makecell{Segment-level\\UAR [\%]}} & \textbf{\makecell{File-level\\UAR [\%]}} & 
    \textbf{\makecell{Segment-level\\UAR [\%]}} & \textbf{\makecell{File-level\\UAR [\%]}}\\
    \midrule
        \textbf{\ac{egemaps}} & 59.0 (5.8) & 58.9 (0.7) & 85.8 (0.6) & 78.2 (0.6) & 58.9 (4.9) & 59.2 (3.5) & 48.8 (2.1) & 51.4 (4.3) \\
        \textbf{\ac{compare}} & 69.4 (6.5) & 68.6 (3.1) & 87.1 (1.1) & 78.6 (2.1) & 67.9 (0.2) & 65.9 (2.0) & 57.0 (1.1) & 57.7 (3.1) \\
        \textbf{\wav} & \textbf{71.2 (10.3)} & \textbf{74.2 (4.9)} & \textbf{97.5 (0.6)} & \textbf{95.7 (1.3)} & \textbf{71.9 (2.3)} & \textbf{69.9 (2.8)} & \textbf{62.7 (8.6)} & \textbf{66.5 (1.8)} \\
    \bottomrule
    \end{tabular}
    }
\end{table}

\begin{table}[t]
    \centering
    
    \scriptsize
    \caption{
    UAR and 95\,\% bootstrap confidence intervals for the 1\textsuperscript{st} fold of the speaker-disjoint partitioning strategy.
    }
    \label{tab:fold1}
    \begin{tabular}{c|cc}
        \toprule
        \textbf{Audio representation} & \textbf{\makecell{Segment-level\\UAR[\%]}} & \textbf{\makecell{File-level\\UAR[\%]}} \\
        \midrule
        \textbf{\ac{egemaps}} & 51.1 (49.3--52.9) & 59.8 (50.8--69.4)\\
        \textbf{\ac{compare}} & \textbf{60.4 (58.7--62.0)} & 71.0 (60.4--80.9)\\
        \textbf{\wav} & 56.6 (54.8--58.3) & \textbf{72.7 (61.6--83.5)} \\
        \bottomrule
    \end{tabular}
\end{table}

Finally, \cref{tab:fold1} provides \ac{UAR} and 95\,\% confidence intervals (CIs) obtained with 1000-sample bootstrapping (with replacement) for the 1\textsuperscript{st} fold of the speaker-disjoint partitioning.
As discussed in \cref{subsec:partitioning}, these results are only provided for comparison purposes with (future) approaches that are too intensive to run on all folds.
Here, the \ac{compare} set gives the best performance for segment-level evaluations while {\wavxlsr} does for file-level ones; however, the CIs show a big overlap in both cases indicating that the differences are not relevant.
In contrast, the \ac{egemaps} lags considerably behind in both cases.
We note that the CIs for segment-level evaluations have a smaller range than the CIs for file-level evaluations, because the sample sizes differ (there are a lot more segments than files).

\section{Discussion}
\label{sec:discussion}

Our findings confirm that COVID-19 detection from free speech samples is basically feasible.
Using standard procedures, we are able to obtain detection rates in the range of 70\,\% \ac{UAR} for the most realistic testing scenarios -- which are comparable to those achieved for other datasets with free speech.
For example, \citet{schuller2021interspeech} report the best result for the COVID-19 speech sub-task of the 2021 \ac{compare} (the baseline) at 72\,\% \ac{UAR}, whereas \citet{dang2022covid} report an AUC of 66\,\% on \textit{CambridgeLongitudinal}, which increases to 79\,\% when utilising the longitudinal nature of the data.
This showcases that speech is suitable for the detection of \ac{COVID}.
Free speech is easier to obtain than read texts/scripted speech, as it can be unobtrusively collected by means of passive recordings instead of requiring the speaker to actively record data and follow any instructions; we therefore expect this type of data to prove highly relevant for future COVID-19 speech research.

An undesirable side-effect of the `wildness' of our data is that it makes the task harder to solve with interpretable features.
This is reflected both in the lower detection performance obtained with \ac{egemaps} and the fact that our acoustic analysis did not yield that clear trends as seen, \eg in studies using data acquired in a more controlled setting, such as~\citet{bartl2021voice}.
Nevertheless, there is some consistency between our present findings and those of \citet{bartl2021voice}: Both studies report shimmer and spectral flux to be relevant for the identification of speakers with \ac{COVID}.
The fact that other features, such as variations in the fundamental frequency or the harmonics-to-noise ratio are not found to be relevant in our study might be related to the different sound type used -- free speech in the present study vs sustained vowels in \citet{bartl2021voice}. 
Moreover, it needs to be considered that the \ac{COVYT} dataset used in the present study includes multiple languages and, thus, a huge range of phonemes and phoneme transitions, making a feature comparison with monolingual studies difficult. 

\section{Contributions and limitations}
\label{subsec:strengthsWeaknesses}
The \ac{COVYT} dataset is the first of its kind \ac{COVID} speech dataset sourced from public multimedia platforms -- a heretofore untapped resource for such data.
Moreover, the presence of the same speaker with and without infection makes it a natural candidate for personalisation approaches, which are expected to improve \ac{COVID} detection performance as previous work found its symptom manifestation to have an individualised component~\citep{dang2022covid}.
These two aspects, combined with the open-source nature of the data, make the \ac{COVYT} dataset a prime basis for further research in voice-based \ac{COVID} detection, which has a huge potential to massively increase future testing capacities while saving waste at the same time.
However, the \ac{COVYT} dataset inherently comes with a number of limitations, especially as the exploited video clips were not recording with the intention to generate data for later scientific analyses. We thus return to the `seven grains of salt' of \citet{coppock2021covid}, and attempt to position the \textbf{contributions} and \textbf{limitations} of our work accordingly.
\begin{enumerate}
    \item \textit{\ac{COVID} vs other diseases}. The \ac{COVYT} dataset does not fulfill this requirement as it primarily contains healthy or COVID-19 positive samples. Information about potential chronic diseases, such as asthma, are not reliably available. However, this also holds true for most previous works. $\rightarrow$~\textbf{Minor limitation}.
    \item \textit{Background noise}. Due to our strict data collection and preparation protocols, we expect only a limited amount of background noise (if at all) to be present in our processed utterances.
    Furthermore, by collecting data `in-the-wild' from several sources, we cover a wide gamut of recording conditions relevant for potential future test applications while still ensuring a higher data quality as compared to fully crowd-sourced datasets. 
    On the downside, most \ac{COVID} positive samples were recorded in isolated environments without other people present, as subjects were under quarantine, whereas \ac{COVID} negative samples encompass a wider gamut of recording environments; see data statistics in \cref{sec:facts}.
    Thus, we are aware of potential systematic differences in background noise and room acoustics/reverberation as well as recording setting, which models may be able to exploit. $\rightarrow$~\textbf{Balanced}.
    \item \textit{Subject knowledge of infection status}. Subjects were not only aware of their infection status, but in many cases created the recordings ad hoc to convey this status to a wider public. 
    Given that our dataset consists of celebrities who rely on building emotional ties with their audience, it is highly possible that some of them modulated their voice accordingly.
    Thus, (intended) emotion could be a potential confounder when building \ac{COVID} detection models on the basis of the \ac{COVYT} dataset.
    Nevertheless, (negative) affect is also a potential disease indicator~\citep{larsen1991day}. 
    Furthermore, it is possible that the vocabulary speakers use when infected (T+) and are making a press-release/interview contains explicit mentions to their health status which can act as `shortcuts' for the models to learn instead of the actual task.
    This is particularly relevant for large, pre-trained models like {\wav} which are known to rely on linguistic information (when available)~\citep{wagner2022dawn, triantafyllopoulos2022probing, shah2021all}. 
    However, this effect is stronger in the deeper transformer layers~\citep{triantafyllopoulos2022probing, shah2021all}, not the earlier convolution ones whence we extracted our embeddings here -- thus we expect this effect to be absent from our study. $\rightarrow$~\textbf{Major limitation}.
    \item \textit{Validity of labels}. Although the labels used here are essentially self-reported, the high level of scrutiny which celebrities are being subject to (especially \wrt a positive diagnosis in the early days of the pandemic) strengthens our confidence in label validity. More problematic than the fact of just knowing a speaker's \ac{COVID} status in terms of negative vs positive is the missing knowledge about (i) the period between a positive \ac{COVID} test and the time of recording, (ii) the type of the used \ac{COVID} test, (iii) the cycle threshold (CT) value at the time of recording in case of a PCR test, (iv) the specific \ac{COVID} variant, (v) the range and severity of the speaker's symptoms at the time of recording, (vi) potential diagnoses of other (chronic) diseases, (vii) the speaker's vaccination status, etc. $\rightarrow$~\textbf{Minor limitation}.
    \item \textit{Data and code availability}. The \ac{COVYT} dataset as well as the code for all experiments presented in this work are publicly released (see~\cref{sec:introduction}). 
    $\rightarrow$~\textbf{Major contribution}.
    \item \textit{Demographic variability}. The \ac{COVYT} dataset does not cover subjects from a wide range of socioeconomic backgrounds; celebrities typically come from the upper echelons.
    Nevertheless, we provide samples in $9$ different languages, produced by speakers from different ethnicities (not all English speakers were natives) and age groups. $\rightarrow$~\textbf{Minor contribution}.
    \item \textit{Speaker-disjoint experiments}. Speaker identity is available for all utterances. Thus, speaker-disjoint partitions can be created. We provide a baseline partitioning scheme to allow standardised evaluation protocols.
    Furthermore, the \ac{COVYT} dataset is the only dataset alongside the \textit{Cambridge Longitudinal} dataset~\citep{dang2022covid}, which contains data of the same speakers with and without infection, which enables future research in personalised approaches that account for individual differences in the manifestation of COVID-19 in human voices.
    $\rightarrow$~\textbf{Major contribution}.
\end{enumerate}

\section{Conclusion}
\label{sec:conclusion}

We introduced the \ac{COVYT} speech dataset for the investigation of (i) the acoustic manifestation of a \ac{COVID} infection as well as (ii) the audio-based automatic detection of COVID-19 in free speech samples. The dataset contains 8+ hours of publicly available audio material and, in contrast to most other datasets in this research field, it features both \ac{COVID} positive and negative speech samples of all 65 included speakers. In our baseline experiments, we identified three acoustic features -- two related to spectral flux and one related to local shimmer -- to differ between the \ac{COVID} positive and negative samples. Moreover, we obtained a \ac{UAR} near 70\,\% for the automatic classification of speech samples according to \ac{COVID} status by using pre-trained speech models. 

The \ac{COVYT} dataset together with the provided benchmarks shall boost further research in the field of speech-based \ac{COVID} detection while ensuring reproducibility and comparability of results. Furthermore, as the dataset contains samples of the same speakers with and without \ac{COVID} infection, we expect it to prove a valuable conduit for future efforts in personalisation approaches that can adapt to the characteristics of individual speakers and, thus, improve performance and reliability.

\section{Acknowledgements}
This work has received funding from the DFG's Reinhart Koselleck project No.\ 442218748 (AUDI0NOMOUS) and from the EU’s Horizon 2020 grant agreement No.\ 826506 (sustAGE).

\ifarxiv
    \section{\refname}
    \printbibliography[heading=none]
\else
    \bibliographystyle{elsarticle-num-names} 
    \bibliography{refs}
\fi
\end{document}